\documentclass{ws-ijmpe}

\def\mytitle{Particle Identification Studies with an ALICE Test TPC}
\def\necon{$\text{Ne-CO}_2\text{-N}_2$}

\begin{document}
  
\markboth{P. Christiansen for the ALICE TPC Collaboration}{\mytitle}

\catchline{}{}{}{}{}
%%%%%%%%%%%%%%%%%%%%%%%%%%%%%%%%%%%%%%%%%%%%%%%%%%%%%%%%%%%%%%%%%%%%

\title{\mytitle}

\author{Peter Christiansen}

\address{Div. of Experimental High-Energy Physics, Lund University,\\
  Box 118, SE-221 00 Lund, Sweden, \\
  peter.christiansen@hep.lu.se}

\author{for the ALICE TPC Collaboration.}

\maketitle

\begin{history}
\received{(received date)}
\revised{(revised date)}
%\accepted{(Day Month Year)}
%\comby{(xxxxxxxxxx)}
\end{history}

\begin{abstract}
  Using a test TPC, consisting of the ALICE TPC field cage prototype
  in combination with the final ALICE TPC readout and electronics, the
  energy loss distribution and resolution were measured for identified
  protons. The measurements were compared to theoretical calculations
  and good quantitative agreement was found when detector effects were
  taken into account. The implications for particle identification are
  discussed.
\end{abstract}

\section{Introduction}
\label{sec:introduction}

The ALICE Time Projection Chamber (TPC) is the main tracking detector
in the central region of the ALICE heavy ion experiment at
LHC~\cite{tpctdr}. In addition to excellent tracking capabilities, the
TPC has particle identification (PID) capabilities through the
measurement of the ionization. This is especially interesting in the
region of the relativistic rise, $\beta\gamma > 3$ ($p_T>3$~GeV/c),
where other ALICE detectors have little or no PID capabilities for
pions, kaons, and protons. STAR has successfully used TPC PID on the
relativistic rise to measure important features of heavy ion
collisions in the regime of intermediate and high
$p_T$~\cite{Abelev:2006jr}.

For PID with TPCs a major limitation is the calibration. So far much
of the PID analysis has used phenomenological descriptions of the
data, based on earlier experience rather than theoretical
calculations. In this paper we compare results from a beam test with a
theoretical model of the gas energy loss and comment on the advantage
of such a study.

\section{Model Description}
\label{sec:model}

In the model of Hans Bichsel~\cite{Bichsel:2006cs}, the calculation of
the energy loss is based on the Fermi Virtual Photon concept,
implemented by Allison and Cobb and elaborated by Bichsel. The
straggling function (Landau distribution), $f$, for the energy loss,
$\Delta$, can then be determined for all track segments, $x$, and
$\beta\gamma$ ($f(\Delta) = f(\Delta,\beta\gamma, x)$).

In general one has the bookkeeping problem that straggling functions
are not analytical functions and for different $\beta\gamma$ and $x$
they are not related by a simple one parameter scaling.  However, it
was shown in the paper (~\cite{Bichsel:2006cs}, Section~8) that two
scaling parameters, $a(\beta\gamma)$ and $b(\beta\gamma)$, can be
found so that: \\ $f(\Delta, \beta\gamma, x) \approx
g(a(\beta\gamma)\cdot \Delta + b(\beta\gamma, x))$. \\ In that way all
straggling functions (and associated truncated mean distributions) can
be approximated by one straggling function (here denoted $g$ for
generic) and a table of two parameter scaling
variables\footnote{Ignoring the high energy tail of the straggling
function one could possibly to a high degree of precision (see
Figure~9 and 10 in~\cite{Bichsel:2006cs}) use only one relevant
parameter, the number of primary collisions $Np = dNp/dx(\beta\gamma)
\cdot \Delta x$.}.

The ionization measured on the TPC pads is related to the energy lost
by the charged particle as it traverses the gas, but there are in
principle six steps -- energy loss, energy deposition, ionization,
electron transport, amplification, and AD conversion
(see~\cite{Bichsel:2006cs} p.~159) -- which should be considered in
going from energy loss of particles to the ADC output measured in the
TPC. So one might ask to what extend the conclusions for the gas
energy loss are also true for the experimental data:

\begin{romanlist}[(ii)]

\item Does the energy loss spectra calculated in the model describe
the measured charge spectra?

\item Can all relevant quantities for PID (such as the truncated mean
distribution) be derived from the straggling function?

\item Is two parameter scaling applicable for experimental
distributions?

\end{romanlist}

\section{Experimental Results}
\label{sec:data}

The results presented here were obtained with a test setup described
in~\cite{Antonczyk:2006qi}, where details of the analysis, and more
results on electronics performance and spatial resolution, can also be
found.

The data were collected with an ALICE Inner Read-Out Chamber (IROC)
which has 63 pad rows with pad length of 7.5~mm each, resulting in a
total track length of 472.5~mm (for the ALICE TPC there is also an
Outer Read-Out Chamber (OROC)). The test TPC was operated with the
newly proposed gas mixture
\necon~(85.7-9.5-4.8)~\cite{Garabatos:2004iv}.

The tracks originated from a secondary beam of single particles from
the CERN PS where the momentum, $p$, of the beam was adjustable
between 1 and 7~GeV/c with a resolution $\sigma_p/p~\sim~0.01$. The
PID was obtained by a TOF setup that allowed the complete separation
of pions and protons up to $p=3$~GeV/c. For each pad row a cluster was
reconstructed and the cluster charge, $Q$, was determined as the sum
of the signals in the cluster.

From the tracks the cluster charge straggling function, $f(Q)$, was
measured, and the ``truncated mean'' distribution, $C$, was derived
track by track from the average of the 60\% lowest cluster charges.

\begin{figure}[htbp]
  \centering
  \includegraphics[keepaspectratio, width=0.45\columnwidth]{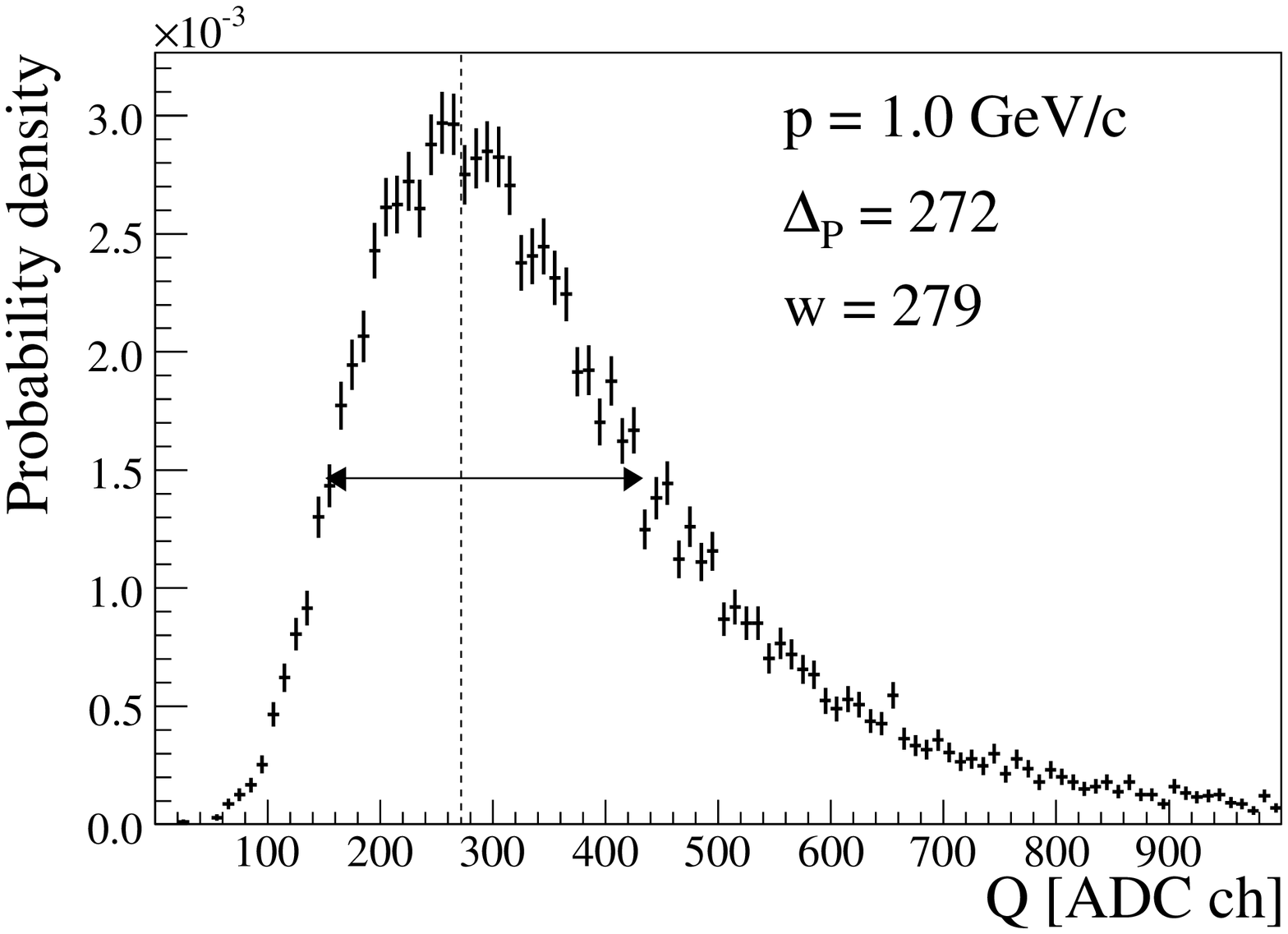}
  \includegraphics[keepaspectratio, width=0.45\columnwidth]{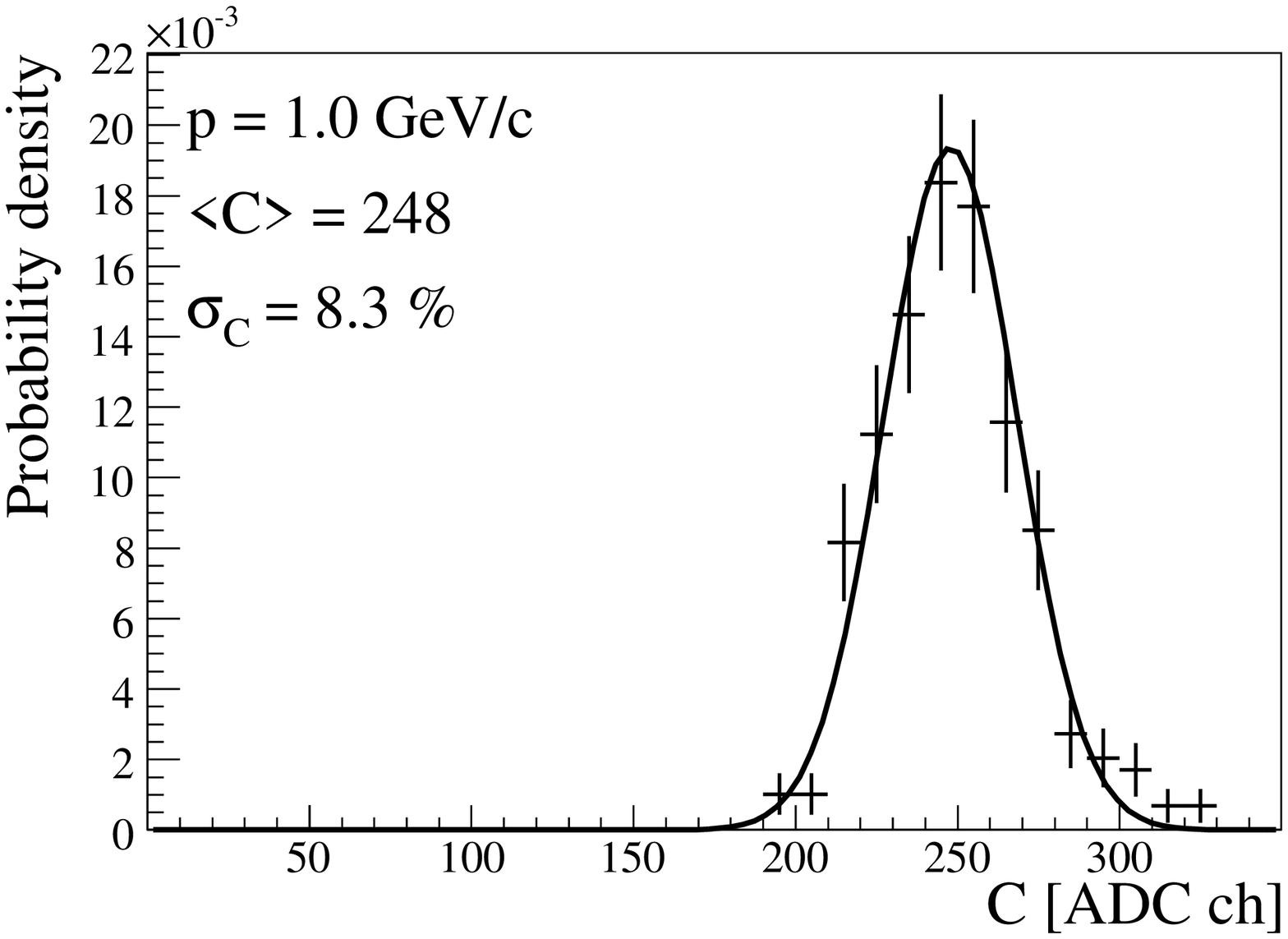}
  \includegraphics[keepaspectratio, width=0.45\columnwidth]{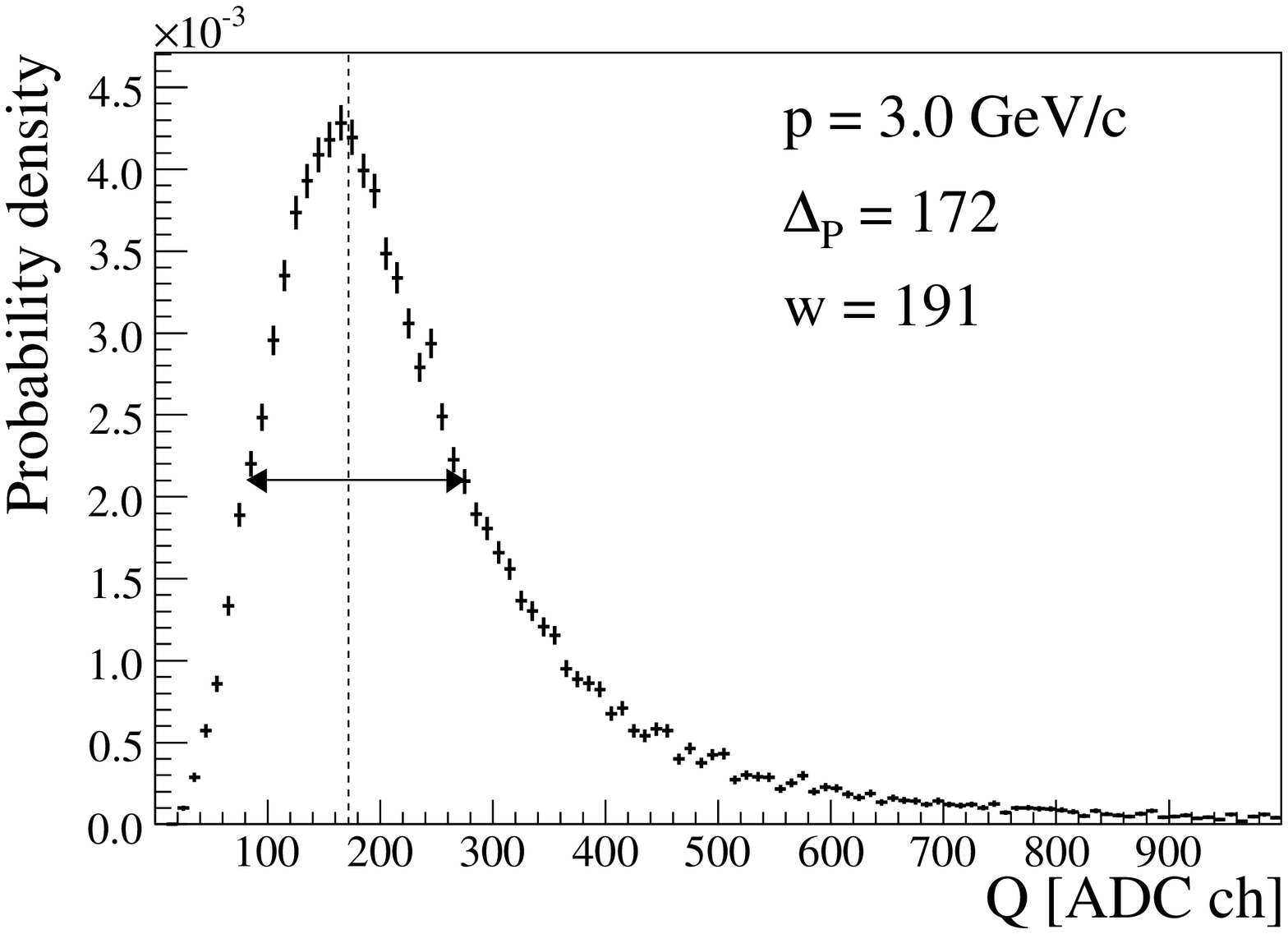}
  \includegraphics[keepaspectratio, width=0.45\columnwidth]{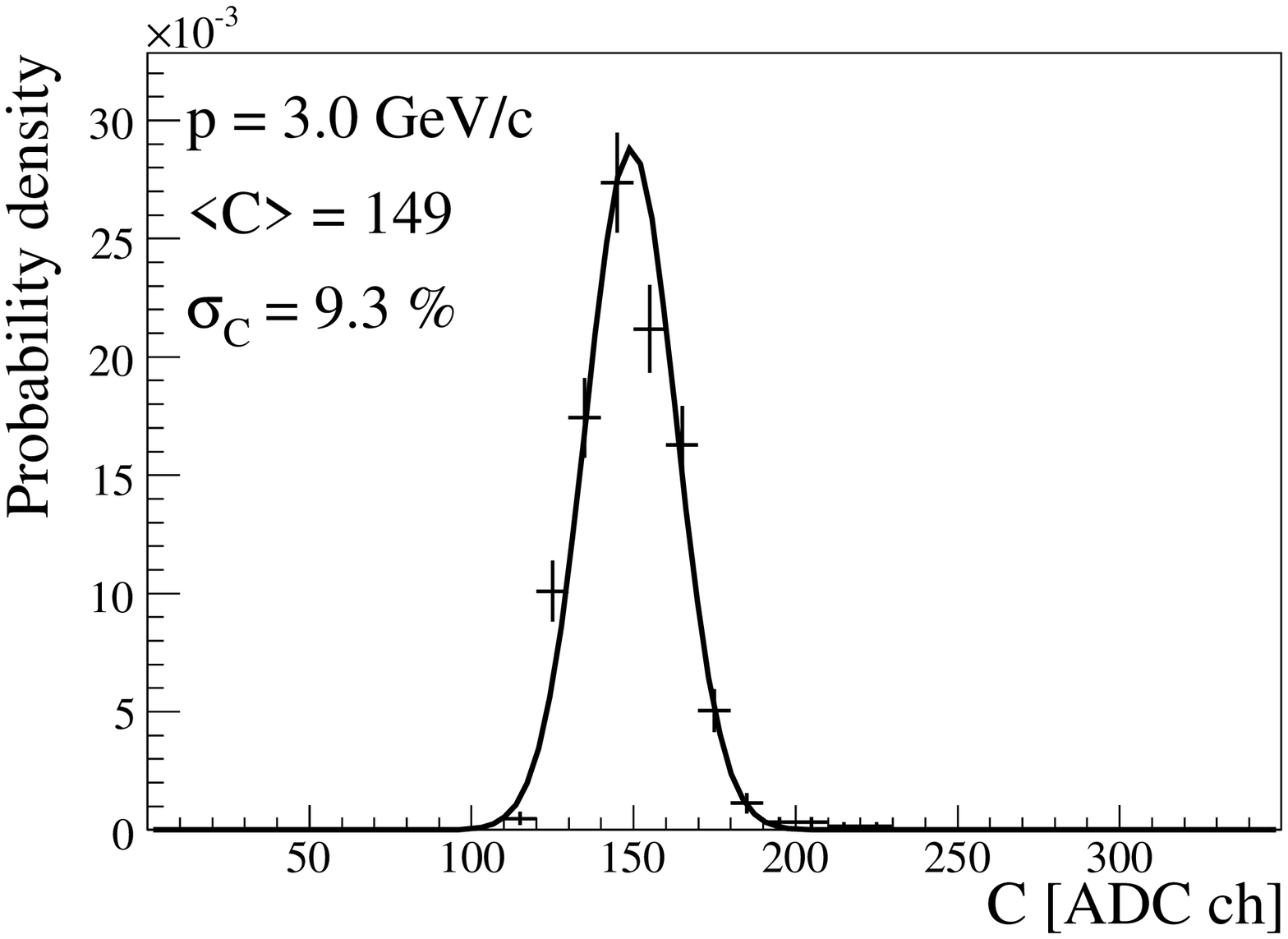}
  \caption{The straggling data (left) and the truncated mean
    distribution (right) for protons with momentum 1 and 3
    GeV/c. Left:~The dashed line indicates the most probable value,
    $\Delta_p$, of the straggling functions while the double arrows
    denote the full width at half maximum FWHM, $w$. Right:~The solid
    line is a normalized Gaussian fit to the distribution.}
  \label{fig:straggling_trunc}
\end{figure}

In the following we study protons with $p=1~\text{GeV/c}$ (294
tracks), and $p=3~\text{GeV/c}$ (614 tracks). This is a small subset
of the beam test data with clean PID. The straggling data and the
truncated mean distributions are shown in
Figure~\ref{fig:straggling_trunc}. The probability of a true Gaussian
fit to the two truncated mean distributions corresponding to $p=1$ and
$3~\text{GeV/c}$ was found to be only 15~\%, and 0.04~\%,
respectively, indicating that the truncated mean is not strictly
Gaussian distributed, with a shoulder at higher $C$ clearly visible,
especially in the 1~GeV/c data.

The energy loss resolution was found to be in agreement with that
determined in the ALICE TPC Technical Design Report~\cite{tpctdr},
when one takes into account that the track length in the full ALICE
TPC is 3.3 times longer (IROC+OROC) than in the prototype (IROC).

\section{Comparison with Model Calculations}

In this section we use the measurements shown in
Figure~\ref{fig:straggling_trunc} and model calculations supplied by
Hans Bichsel to study the questions outlined in
section~\ref{sec:model}.

\begin{figure}[htbp]
  \centering
  \includegraphics[keepaspectratio, width=0.45\columnwidth]{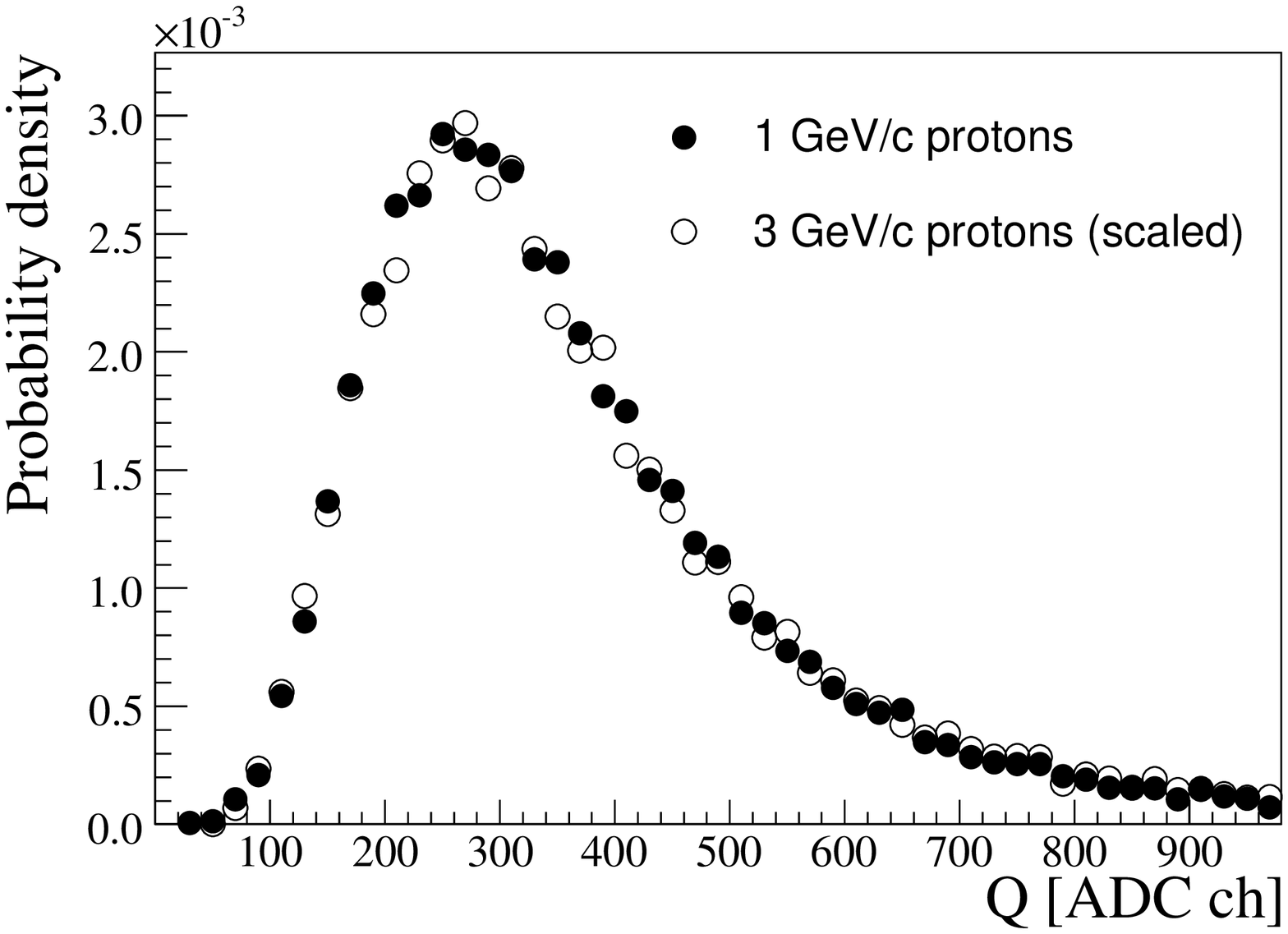}
  \includegraphics[keepaspectratio, width=0.45\columnwidth]{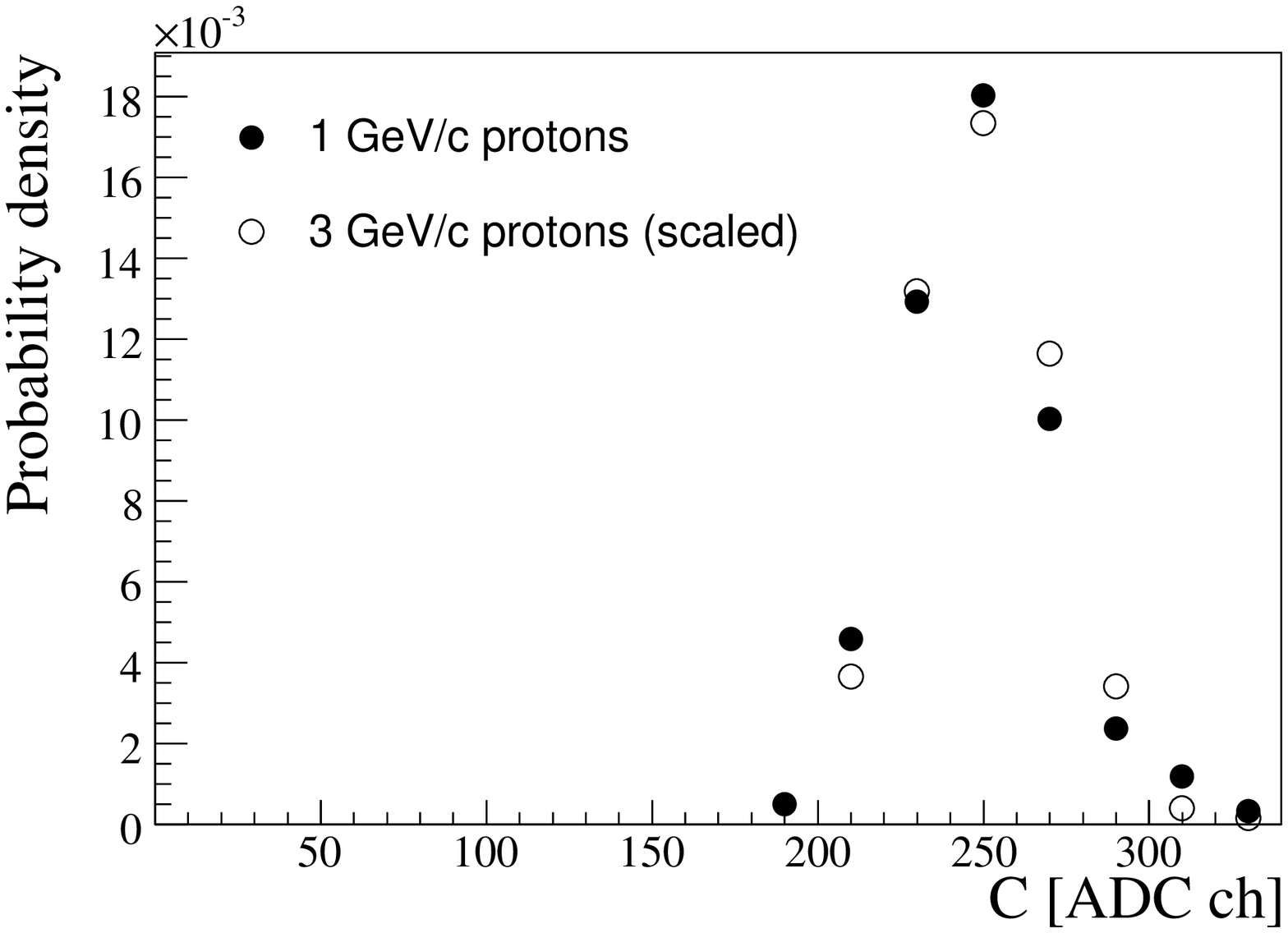}
  \caption{The 3 GeV/c proton straggling data (left) and truncated
  mean distribution (right) scaled to match the 1 GeV/c
  distribution. The scaling parameters are $a = 1.46$, and $b = 33$,
  see text. Note that the same set of scaling variables is used for
  both comparisons.}  
  \label{fig:scaling}
\end{figure}

Figure~\ref{fig:scaling} shows an overlap of the measured straggling
data for $1~\text{GeV/c}$ protons and the \emph{scaled} measured
straggling data for $3~\text{GeV/c}$ protons (left), and similarly the
overlap of the two truncated mean distributions (right). The scaling
parameters were determined from the data. The close agreement
demonstrates the applicability of two-parameter scaling between
Minimum Ionizing Particles (MIPs) ($p=3$~GeV/c protons) and particles
with an energy loss similar to that on the plateau ($p=1$~GeV/c
protons has similar energy loss spectra as 1000~GeV/c protons (except
for the large $\Delta$ tail)).  Since the scaling works for the
smallest length scale (the pads in the OROCs are longer: 10~mm and
15~mm) and for the relevant energy loss range for high-$p_t$ PID, this
scaling is probably applicable to all relevant distributions. \\

\begin{figure}[htbp]
  \centering
  \includegraphics[keepaspectratio,width=0.45\columnwidth]{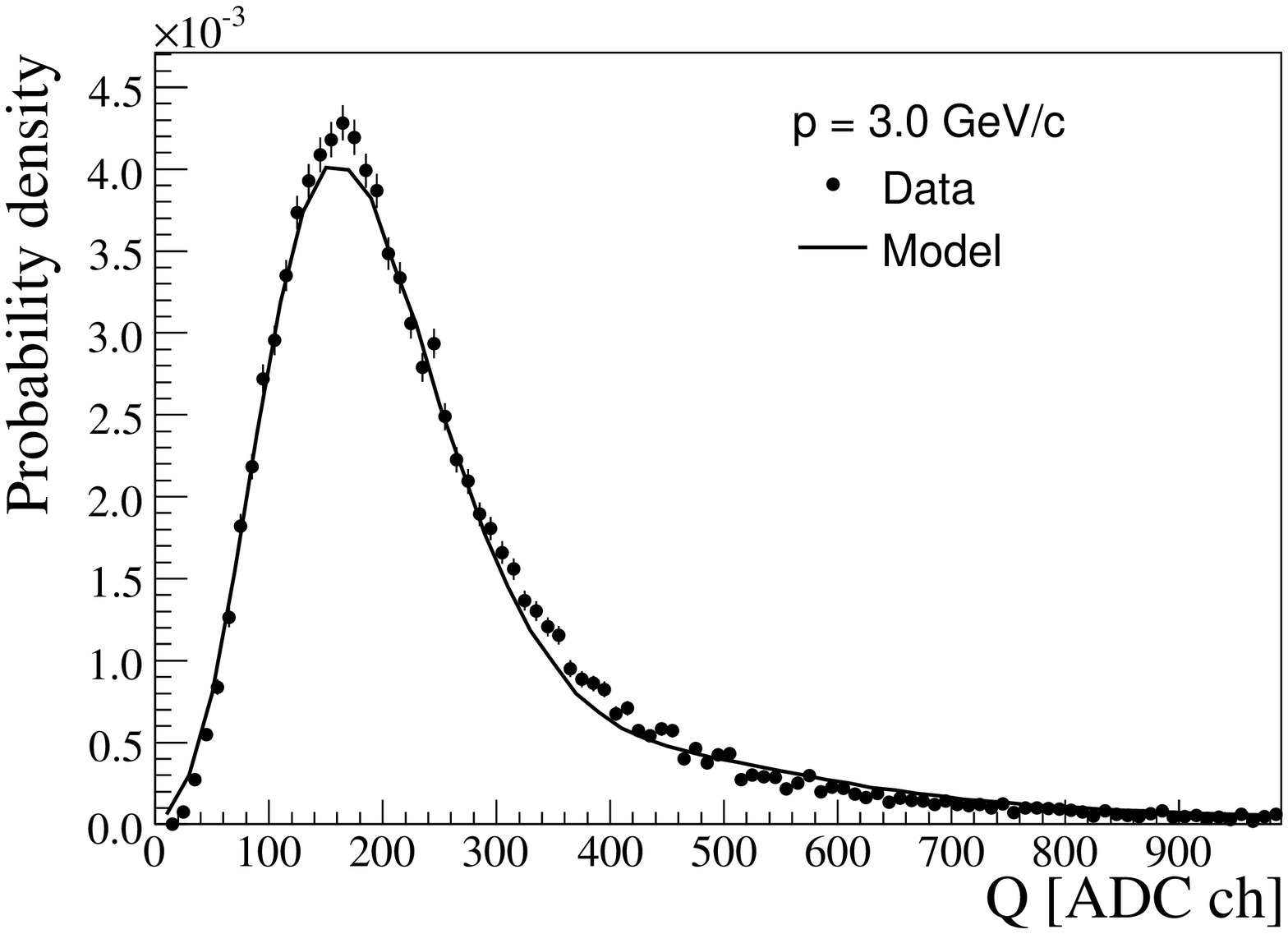}
  \includegraphics[keepaspectratio,width=0.45\columnwidth]{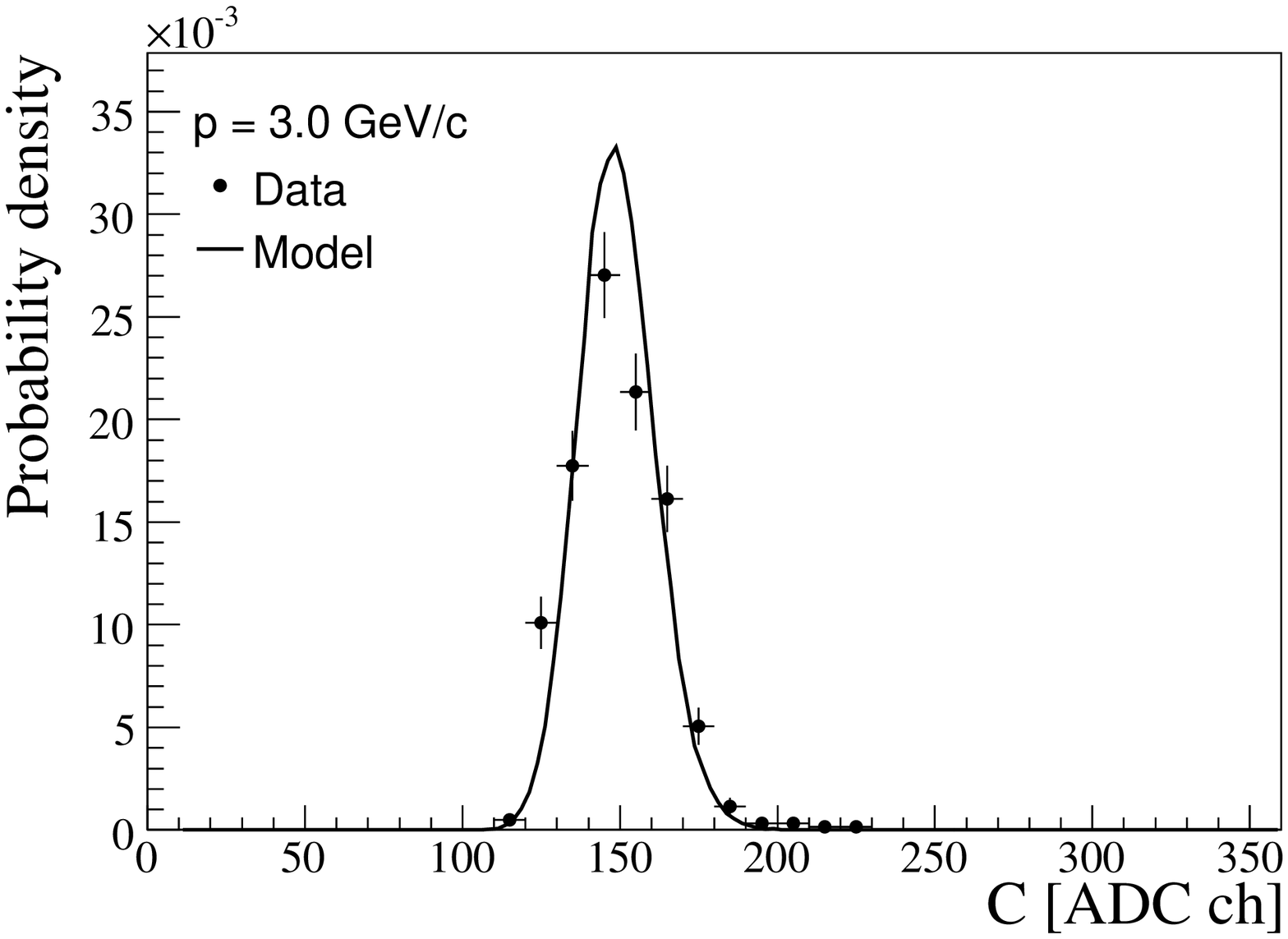}
  \caption{Comparison of the measured cluster charge straggling data
  and truncated mean distributions (data) to the energy loss functions
  calculated in the Bichsel model (solid line).}
  \label{fig:first_comparison}
\end{figure}

Figure~\ref{fig:first_comparison} shows a comparison between the
experimental cluster charge data and \emph{energy loss} calculations
for 3~GeV/c protons. The conversion factor (ADC~ch/eV) is the only
free parameter and has been adjusted to align the peaks; the overall
normalization is fixed by requiring the integral to be unity. The
agreement between the two straggling functions is reasonable, but the
extracted resolution from the truncated mean is $\sigma_C= 8.0~\%$ in
the model and 9.3\% for the data (Figure~\ref{fig:straggling_trunc}),
leaving a discrepancy of $15~\%$.

This difference is also found when comparing the experimental
straggling data to the experimental truncated mean distribution. If
cluster charges are randomly generated with probabilities according to
the experimental straggling data (shown in
Figure~\ref{fig:straggling_trunc}) and a virtual track is constructed
with the same number of clusters as for the data, the resolution is
close to the 8 \%, in agreement with the model, but not with the
experimental results. The resolution of the real tracks is
deteriorated because cluster charges are correlated; when cluster
charges in neighboring pad rows are compared they exhibit a $+33 \%$
correlation factor, which reduces the generic information for the
track.

\begin{figure}[htbp]
  \centering
  \includegraphics[keepaspectratio, width=0.45\columnwidth]{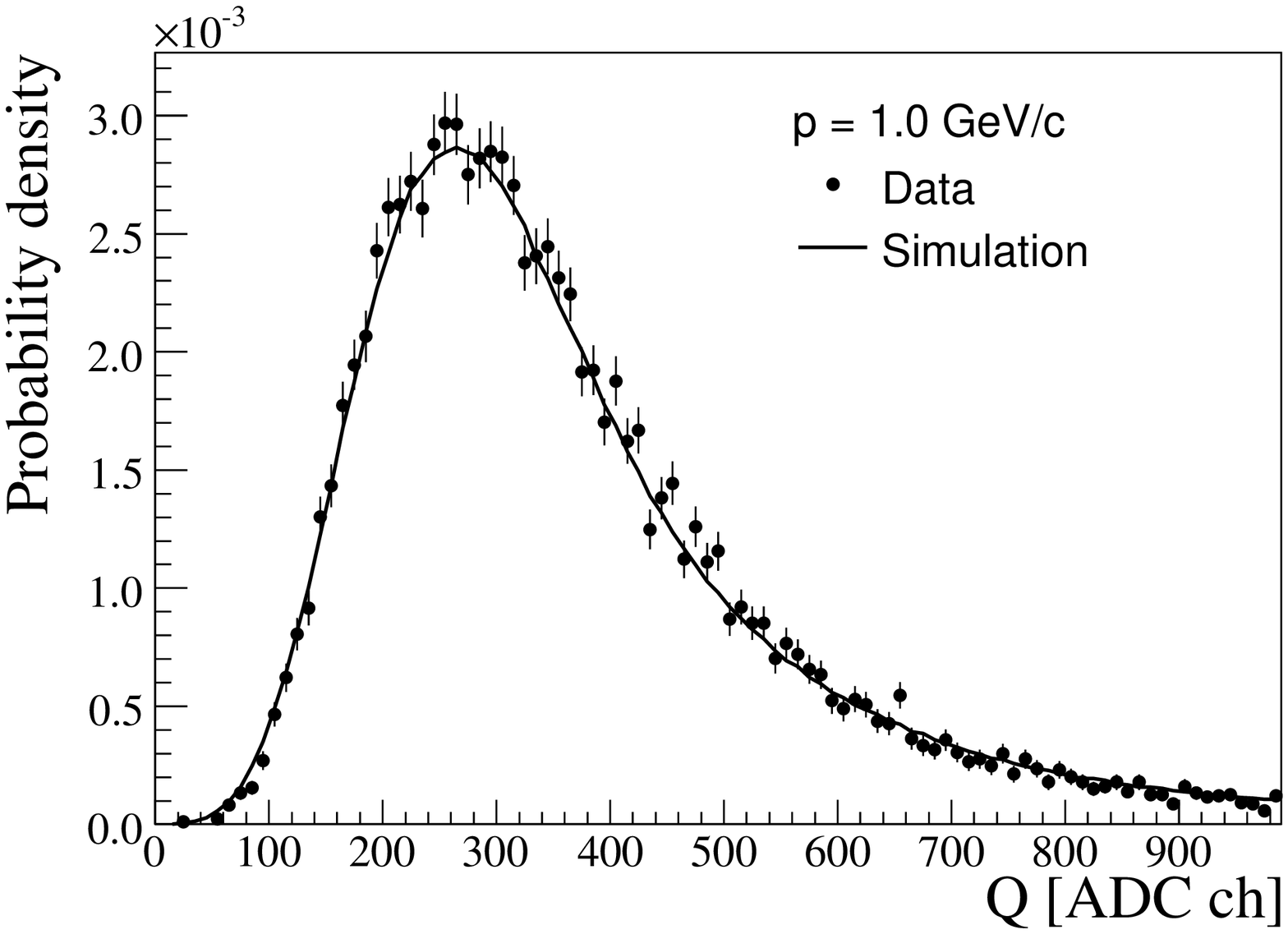}
  \includegraphics[keepaspectratio, width=0.45\columnwidth]{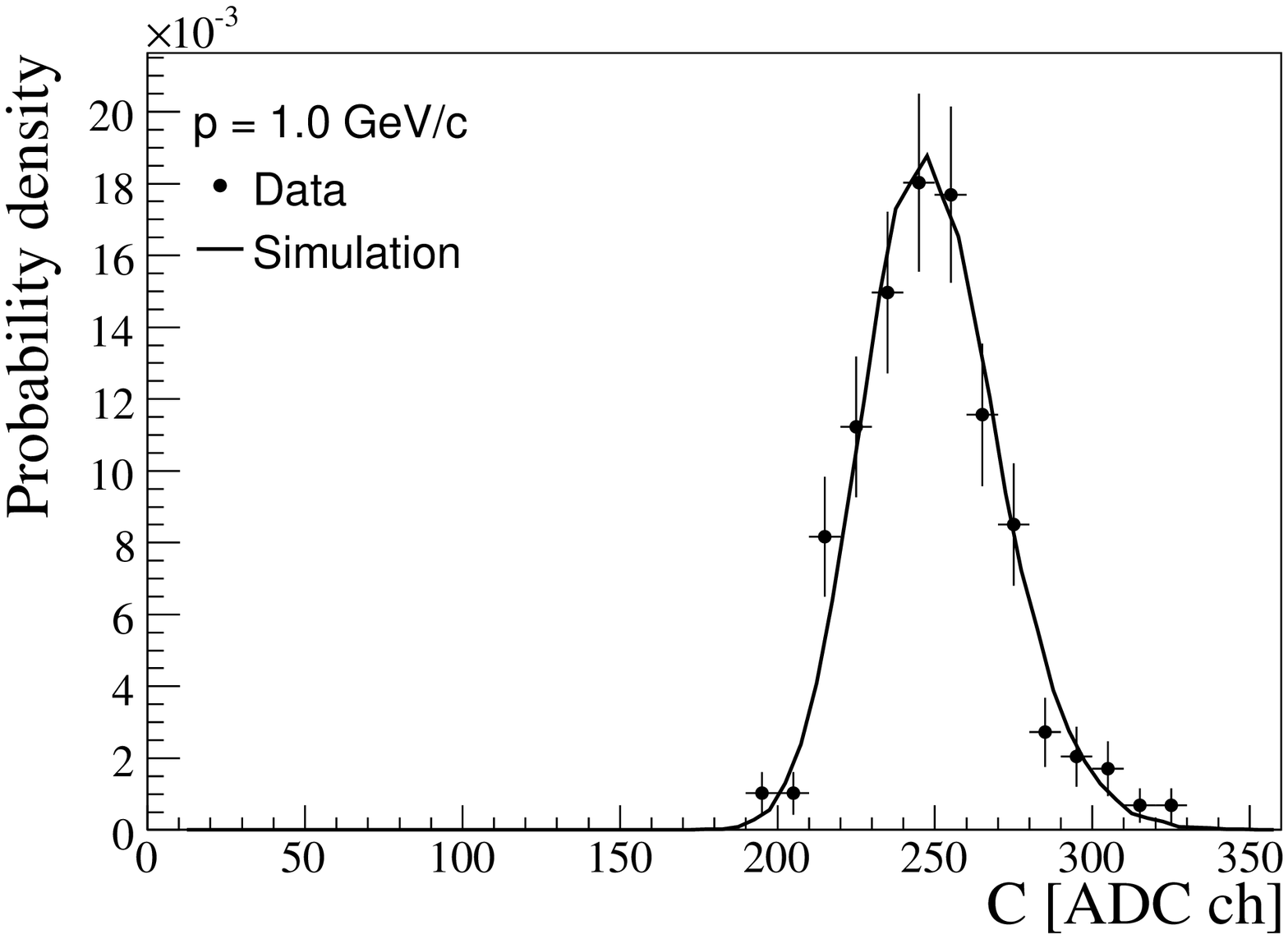}
  \includegraphics[keepaspectratio, width=0.45\columnwidth]{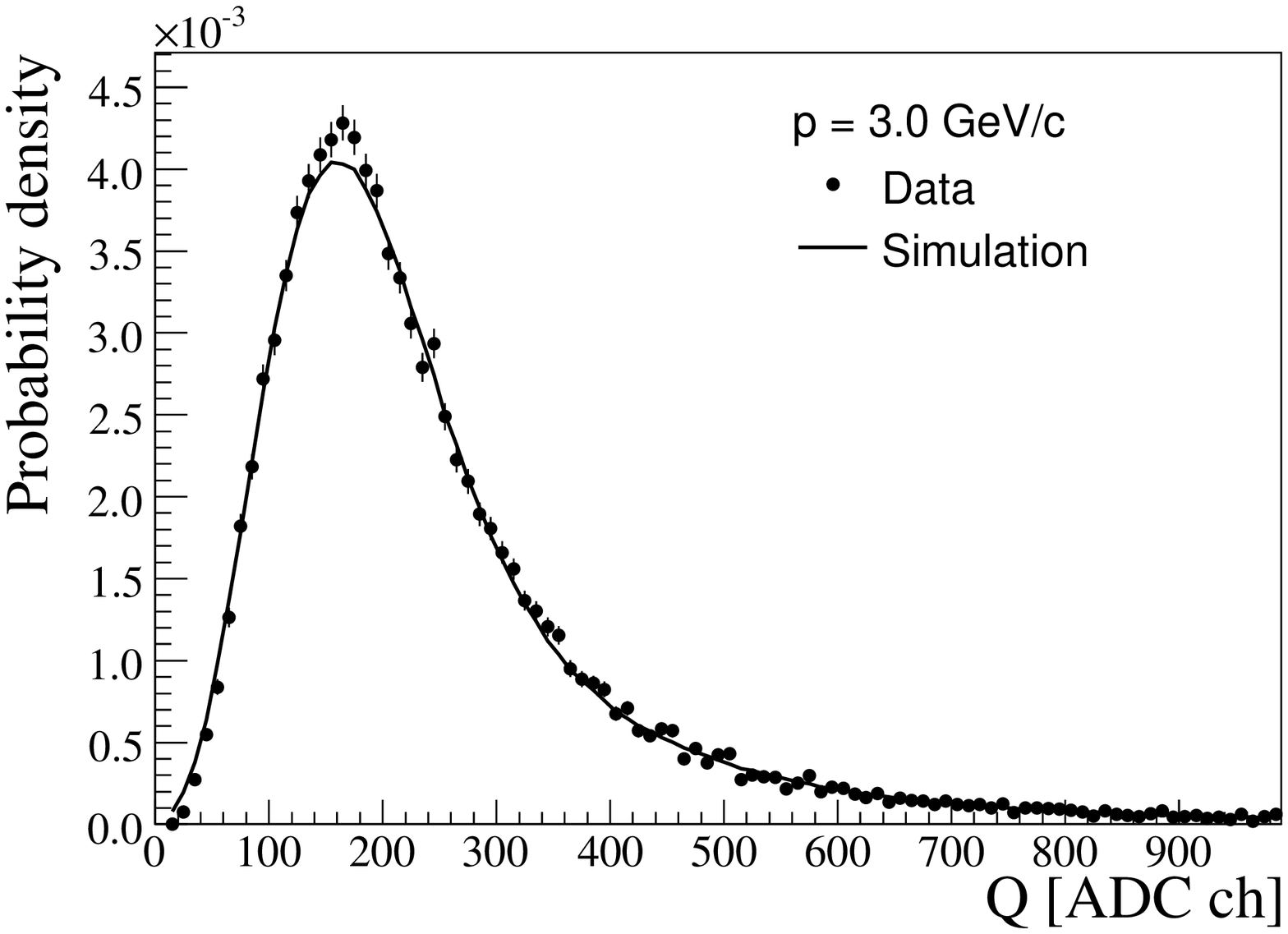}
  \includegraphics[keepaspectratio, width=0.45\columnwidth]{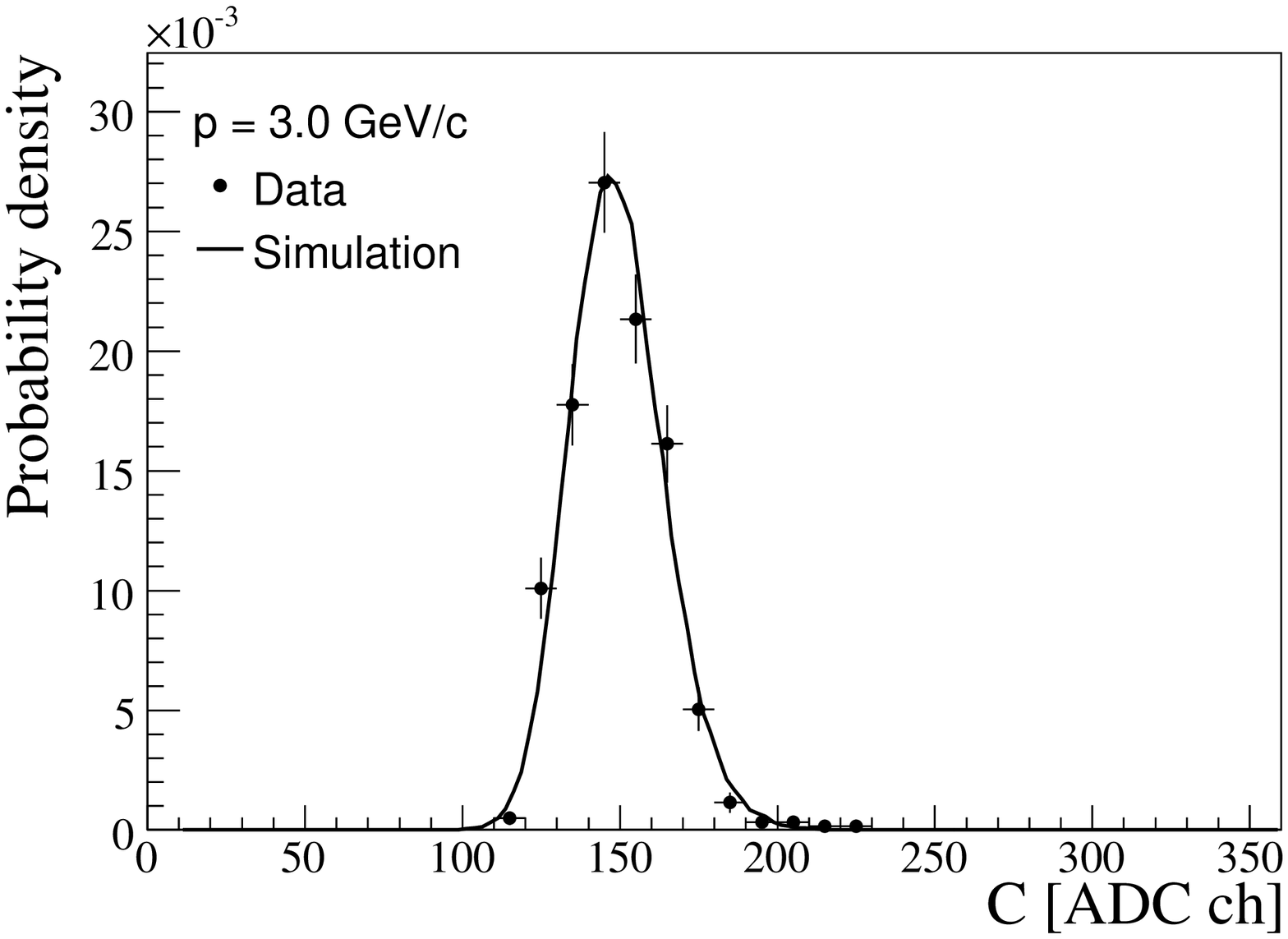}
  \caption{Comparison of the measured straggling data and truncated
  mean distributions (experimental data) to the model of Bichsel with
  detector effects included (simulation).}
  \label{fig:bichsel_comparison}
\end{figure}

The correlations in the cluster data originate from detector effects
during the six steps from energy loss to ADC output mentioned in
section~\ref{sec:model}. Figure~\ref{fig:bichsel_comparison} shows the
comparison between the data and the adjusted Bichsel model modified
for diffusion (electron transport) and exponential gain amplification
variations (amplification), see~\cite{Antonczyk:2006qi} for
details. From the agreement we conclude that the energy loss mechanism
and the two detector effects combined are sufficient to describe the
data.

This method also fixes the adjustment of the gain to $G =
9.6~\text{ADC ch/electron}$ or an effective gas amplification gain of
9~600 (the fraction of the signal picked up by the pads).

If the simulated distributions are treated as fits to the truncated
mean distributions, the probability of the model describing the data
is 21~\%, and 0.11~\%, for the $p=1$ and $3~\text{GeV/c}$ data sets
respectively. In both cases, the description of the data is better
than the previous Gaussian fit functions in
Figure~\ref{fig:straggling_trunc} where, for each setting, there are
two fit parameters to adjust. For the energy loss model the two gas
gains differ by less than 3~\% which could be due to gas density
variations between the two runs.

\section{Conclusion}
\label{sec:conclusion}

We have found that it is possible to describe the test data with
calculations from first principles when detector effects are taken
into account. This good agreement between model and data led Hans
Bichsel to propose an optimization of the ALICE Monte Carlo
simulation~\cite{Bichsel:2006yg}.

The tracks are not fully characterized by their straggling functions
since the cluster charges in a track are correlated. So it is not
possible to directly use energy loss calculations to optimize
PID. Note also that a fit to the cluster charges of a single track to
extract e.g. the most probable energy loss as a PID estimator will
face the same problems since the data are correlated, so that the
fitting assumption of independent data is not fulfilled.

It is interesting to note that the two parameter scaling relation is
still true for the experimental data. This might only be true in the
case, as here, where one has almost identical track geometry, so that
effects of diffusion are the same.

Even though the disagreement between the resolution derived from the
experimental straggling data and the one measured from the truncated
mean could have been found in the data, it was only realized after
comparing to the model calculations, so we would like to stress that a
quantitative precise model of the energy loss in gases is an important
tool for calibrating and understanding a TPC (and any other gas
detector).

\section{Acknowledgments}

The authors would like to thank Hans Bichsel for many valuable
discussions on energy loss in gases, PID, and for supplying us with
model calculations.

\end{document}